\documentclass[12pt]{iopart}
\usepackage{graphicx}
\usepackage{xspace}

\newcommand{\sqrts}{$\sqrt{s_{_{NN}}}$ \xspace}
\newcommand{\npart}{$N_{part}$ \xspace}
\begin{document}

\title[Volume Effects on Strangeness Production]{Volume Effects on Strangeness Production}

\author{Helen Caines for the STAR Collaboration
\footnote[3]{Correspondence address: helen.caines@yale.edu
}\footnote[2] {For full author list and acknowledgements see
Appendix "Collaborations" of this volume}}

\address{WNSL, Physics Dept., Yale University, New Haven, CT 06520, U.S.A.}

\begin{abstract}

 A study of the yields of strange particles produced in
 heavy-ion and elementary collisions is presented using
 preliminary results from the STAR experiment at RHIC.
 The strange particle production rates, relative to those of $p+p$,
 have been proposed as a  means of determining an enhancement of
strangeness production in heavy-ion collisions. Analysis of results
from STAR show that this enhancement measure is reduced, or
comparable, when contrasted to that at top CERN SPS energies. A
smaller suppression in the $p+p$ yields at RHIC energies due to
finite volume effects could be the cause of such a result.
 By studying the yields as a function of centrality we hope to establish
 how these effects vary with the volume of the source created.

\end{abstract}



\vspace{-0.7cm}
\section{Introduction}

\noindent A vast amount of work has been done in implementing
statistical models to aid our understanding of heavy-ion collisions,
\cite{StatModels} and references therein. One of the most important
features of these statistical models is that they assume a thermally
and chemically equilibrated system at chemical freeze-out. They
assume that the system consists of non-interacting hadrons and
resonances but make no predictions about how the system arrived in
such a state, or how long it exists in such fashion. Given these
conditions the number density of a given particle can be calculated
for a given chemical freeze-out temperature, T$_{ch}$,
baryo-chemical potential, $\mu_{B}$, strangeness potential,
$\mu_{s}$, and strangeness saturation factor, $\gamma_{s}$. Another
important issue is that statistical models generally utilize Grand
Canonical Ensemble statistics, which are only appropriate when the
system becomes large.  In small systems, or the (micro)Canonical
regime, all quantum numbers have to be conserved explicitly; this
means there not only has to be
 energy available for strangeness creation but also the phase space.
 This leads to an interesting
effect on strange particle production; one of a suppression of
strangeness in small systems due to a lack of available phase space.
 Once the volume is sufficiently large, this phase space suppression
 disappears and the amount of strange particle creation per unit volume becomes constant.
Statistical models using the Grand Canonical approach may still
appear to work for small systems but the fits do not represent true
temperatures and chemical
  potentials. Hence, we need to establish at what collision energy and
  correlation volume, if any, such a Grand Canonical state occurs.

~

\begin{figure}
 \begin{center}
  \includegraphics[width=0.9\textwidth]{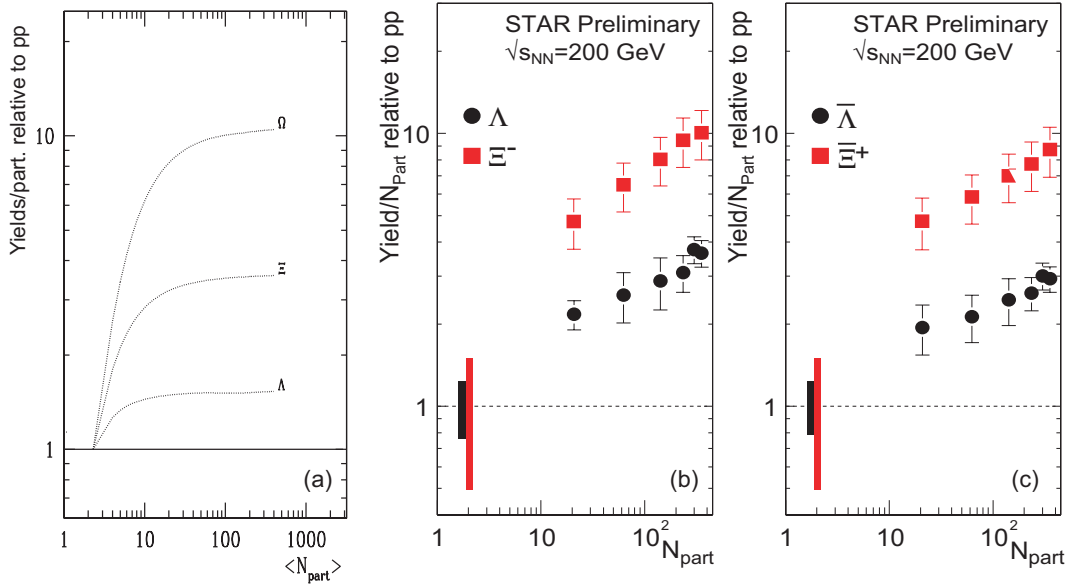}\\
\caption{ (a) Predicted enhancement factors as a function of
particle species for Pb+Pb collisions at \sqrts = 130
GeV~\cite{Redlich}. Preliminary enhancement factors vs of \npart for
(b) $\Lambda$ and $\Xi^{-}$ (c) $\bar{\Lambda}$ and $\bar{\Xi}^{+}$
in Au+Au collisions at \sqrts= 200 GeV. Error bars are statistical.
Ranges for $p+p$ data indicate the systematic
uncertainty.}\label{Fig:Enhance}
\end{center}
\vspace{-0.8cm}
\end{figure}

\noindent  Phase space suppression effects are measured
experimentally as the
  yield per participant relative to the yield per participant in $p+p$ (or $p+Light$ nuclei),
   the volume of the system is believed to be directly
proportional to the number of participants, \npart.
Fig.~\ref{Fig:Enhance}(a) shows the predicted behaviour  for
$\Lambda$ and $\Xi$  in Pb+Pb collisions at \sqrts = 130 GeV as a
function of \npart, or volume~\cite{Redlich}. No calculations are
available for Au+Au data at 200 GeV. As can be seen the larger the
  number of strange quarks in the particle the greater the phase space suppression effect.
  It has also been demonstrated that
  increasing the collision energy decreases the suppression of a given species~\cite{Redlich}.

\subsection{Enhancement Factors}

\noindent At very low energies, such as those measured by the KAOS
experiment at SIS, we can see the effects of canonical suppression
even in the kaons~\cite{Kaos, Oeschler}. As the collision energy
increases this kaon suppression dissipates and it has been shown
that in Pb+Pb collisions of \sqrts = 17.3 GeV even the $\Lambda$
yield per participant appears to saturate~\cite{Bruno}. It would
seem therefore that the top energy CERN collision data show evidence
of the applicability of the Grand Canonical Ensemble for particles
up to the multi-strange baryons. The more recent data at \sqrts =
8.8 GeV~\cite{Bruno} , however, show enhancement factors for the
$\Xi$ and $\Lambda$ that are approximately equal to the 17.3 GeV
data. This result goes against our understanding of how canonical
suppression is related to collision energy. Calculations have shown
that the enhancement for $\Xi$ should be much higher at \sqrts = 8.8
GeV than at \sqrts = 17.3 GeV~\cite{Redlich}.

~

\noindent The predictions of Fig.~\ref{Fig:Enhance}(a) can be
compared to Fig.~\ref{Fig:Enhance}(b)and (c), which shows
preliminary calculations of the enhancement factors for strange
hyperons from STAR~\cite{Starpp, Star200}. The difference in
suppression from 130 to 200 GeV is expected to be small. We see that
for this data set the hyperons show no sign of reaching a plateau,
this could be an over-population of strangeness in the $\Lambda$
channel. However, the $\gamma_{s}$ factor, calculated as a function
of centrality from a statistical model~\cite{PBMStat}, only
approximately reaches unity for the most central data~\cite{Star200,
StarGammaS}. This indicates that the created medium at RHIC is only
just reaching the Grand Canonical Ensemble limit for the most
central collisions. Figure~\ref{Fig:Enhance}(a) predicts the
saturation levels of the different enhancement factors, the measured
results are above those from theory. Fig.~\ref{Fig:Enhance}(b) and
(c) also appear to show a smooth transition from $p+p$ to the most
central Au+Au results, in contrast to the sharp rise at small \npart
predicted by theory. Improved measurements as a function of
centrality will help clarify this issue.

~

\noindent There are several possible explanations for the
discrepancies between the RHIC and SPS data and theory. One is that
the freeze-out conditions of the sources are not those assumed in
the calculations, the enhancement factors being very sensitive to
this assumption. Another possibility is that the correlation volume
is not linearly proportional to \npart.

\section{Scaling Variables}

\begin{figure}
  \begin{center}
 \includegraphics[width=0.9\textwidth]{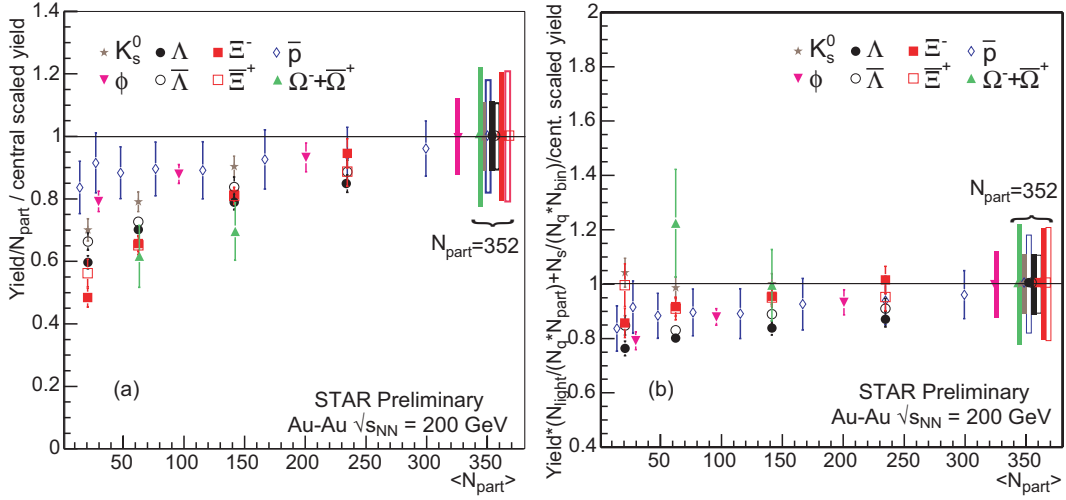}\\
  \caption{(a) Yield per \npart (b) yields scaled as per Eqn.~\ref{Eqn:Scale} for various
   particle species  normalized to the measurement at $N_{part}$ = 352, the most central Au+Au collisions at
200 GeV versus $N_{part}$.}\label{Fig:NScale}
\end{center}
\vspace{-0.8cm}
\end{figure}

\noindent No matter how the transition from phase space suppression
to the Grand Canonical regime occurs in A+A collisions it is evident
that there is an effect. A study of how the particle yields in Au+Au
collisions depend on the centrality of the collision has therefore
been made. Figure~\ref{Fig:NScale}(a) shows the yields of various
particles scaled by \npart relative to the yield per \npart in the
most central Au+Au collisions. It can be seen that the anti-protons,
which contain no strange quarks, scale linearly with \npart. As the
strange quark content of the mesons and baryons increases this
scaling steadily breaks. The  $\Omega$, which contains 3 s quarks,
clearly has no such linear dependence on \npart. This led to the
idea that strange quarks have a different scaling to the light u and
d quarks. One possible scaling is with the total number of binary
collisions, $N_{bin}$, in the collision. The individual particle
scaling therefore becomes:
\begin{equation}\label{Eqn:Scale}
   N_{light}*N_{part}/N_{q} + N_{s}*N_{bin}/N_{q}
\end{equation}

\noindent where $N_{light}$ is the number of u and d quarks in the
particle, $N_{s}$ is the number of strange quarks and $N_{q}$ is the
total number of quarks in the hadron. Thus an anti-proton still
scales with \npart, the $K^{0}_{s}$ now scales as
0.5*$N_{part}+0.5*N_{bin}$ and the $\Omega$ scales as $N_{bin}$. The
results of such as scaling for the Au+Au 200 GeV data is shown in
Fig.~\ref{Fig:NScale}(b), it appears to be quite successful. This
suggests that the relevant volume for strange particle production is
not merely geometrical, and thus controlled by \npart, but is
strongly affected by the number of hard processes in the collision.
It should be noted that the $\phi$ is anomalous and is scaled in
Fig.~\ref{Fig:NScale}(b) by \npart and not by $N_{bin}$. Perhaps
this is indicative that the $\phi$, which contains an $s\bar{s}$,
quark pair, is created via a different mechanism.

\section{Summary}

\noindent In summary we have shown that it is likely that the Grand
Canonical regime is applicable for the most central Au+Au collisions
at RHIC. The data shows that the correlation volume for strange
particle production is not linearly correlated to \npart but appears
to be linked to $N_{bin}$. Further studies are needed to determine
how the correlation volume can be mapped , if at all, onto a
physically measurable quantity and a model developed that can
explain both the SPS and RHIC data. The approach of the 200 GeV
Au+Au data to the Grand Canonical regime does not seem to be well
described by theory. The high statistics Au+Au data, already taken
at 200 GeV, and the upcoming Cu+Cu run will allow us to map out  the
region around \npart $\sim$ 10 in more detail. This region is
critical for testing the parametric dependence of suppression on
volume.

~


\begin{thebibliography}{99}


\bibitem{StatModels} A.~Bialas, (2003) Nucl. Phys.  {\bf A715} 95c,
J.~Rafelski and J.~Letessier,(2003) Nucl. Phys. {\bf A715} 97c ,
V.~Koch (2003) Nucl. Phys. {\bf A715} 108c

\bibitem{Redlich} A.~Tounsi, A.~Mischke and K.~Redlich, (2003) Nucl. Phys. {\bf A715} 565

\bibitem{Kaos} R.~Barth {\it et. al.}, (1997) Phys. Rev. Lett. {\bf 78}  4007

\bibitem{Oeschler} H.~Oeschler J.~Cleymans and K.~Redlich nucl-ex/0112005

\bibitem{Bruno} G.E.~Bruno {\it et al.} (NA57 Collaboration) (2004) J. Phys. G 30  S717

\bibitem{Starpp} M.~Heinz (STAR Collaboration) These proceedings,
R.~Witt (STAR Collaboration) These proceedings

\bibitem{Star200} M.~Estienne (STAR Collaboration) These proceedings.

\bibitem{PBMStat} P.~Braun-Munzinger, I.~Heppe and J.~Stachel, (1999) Phys.
Lett. {\bf B465}, 15

\bibitem{StarGammaS} O.~Barannikova (STAR Collaboration) nucl-ex/0408021

\end{thebibliography}
\end{document}